# Transferable conditional brain charts link cortico-subcortical structure to Alzheimer's disease progression

**Authors:**

Nitin Sharma[1,2], Sicheng Dai[1,3], Francesco A. Mallus[1], Atharva Kand[2], Marc Lemnitzer[1], Minh-Hieu Nguyen[1], Mebin Muhammed Hamzakoya Sulaikha[2], Jijia Xing[2], Mariam Zabihi[1], Gülce Lale[1,4], Gokul Perumbayil Vijayakrishnan[1], Elias Schreiner[1], Madhumitha Sivakumar Punitha[2], Paul Buttkus[5], Çağatay Yıldız[6], Poornima Kumar[7,8], Thomas Grote[9], Esten Leonardsen[10,11], Francesco Sammarco[1], Vinod Kumar Jangir[12], Sarah Sobotta[13], Barbora Rehák Bučková[14,15], Augustijn A. A. de Boer[14,15], Konstantinos Tsilimparis[14,15], Christian Gaser[16,17,18], Charlotte Fraza[14,15,19], Saige Rutherford[20], Seyed Mostafa Kia[14,21,22], Andre F. Marquand[14,15], for the Alzheimer's Disease Neuroimaging Initiative*, the Australian Imaging Biomarkers and Lifestyle flagship study of ageing**, Yanwu Yang[1,3#], Richard Dinga[21#] & Thomas Wolfers[1,2,18#]

**Affiliations:**

1. Department of Psychiatry and Psychotherapy, University Hospital Tübingen, Tübingen, Germany

2. Institute of Psychology, Friedrich Schiller University Jena, Jena, Germany

3. German Center for Mental Health (DZPG), partner site Tübingen, Tübingen, Germany

4. Department of Women's and Children's Health, Science for Life Laboratory, Uppsala University, Uppsala, Sweden

5. Synergy of Systems Group, Center for Interdisciplinary Digital Sciences, Technische Universität Dresden, Dresden, Germany

6. Cluster of Excellence "Machine Learning: New Perspectives for Science", University of Tübingen, Tübingen, Germany

7. Center for Depression, Anxiety and Stress Research, McLean Hospital, Belmont, MA, USA

8. Department of Psychiatry, Harvard Medical School, Boston, MA, USA

9. Department of Computer Science and Artificial Intelligence, University of Technology Nuremberg, Nuremberg, Germany

10. Department of Psychology, University of Oslo, Oslo, Norway

11. Centre for Precision Psychiatry, Oslo University Hospital and Institute of Clinical Medicine, University of Oslo, Oslo, Norway

12. Max Planck Institute for Biological Cybernetics, Tübingen, Germany

13. Lise Meitner Research Group Cognitive Neurogenetics, Max Planck Institute for Human Cognitive and Brain Sciences, Leipzig, Germany

14. Donders Institute for Brain, Cognition and Behaviour, Radboud University, Nijmegen, the Netherlands

15. Department of Cognitive Neuroscience, Radboud University Medical Center, Nijmegen, the Netherlands

16. Department of Psychiatry and Psychotherapy, Jena University Hospital, Jena, Germany

17. Department of Neurology, Jena University Hospital, Jena, Germany

18. German Center for Mental Health (DZPG), partner site Halle-Jena-Magdeburg, Jena, Germany

19. Institute of Psychiatry, Psychology and Neuroscience, King's College London, London, United Kingdom

20. Department of Statistics, Indiana University, Bloomington, IN, USA

21. Department of Cognitive Science and Artificial Intelligence, Tilburg University, Tilburg, the Netherlands

22. Department of Psychiatry, UMC Utrecht Brain Center, University Medical Center Utrecht, Utrecht, the Netherlands

**Keywords:** normative modelling, brain charts, lifespan, cortical thickness, subcortical volume, site transfer, Alzheimer's disease, GAMLSS

***Corresponding authors:***

*Nitin Sharma,*

*Laboratory for Mental Health Mapping (MHM-Lab),*

*Friedrich Schiller University Jena, Jena, Germany,*

*nitinsharma3150@gmail.com*

*Prof. Dr. Thomas Wolfers*

*Carl-Zeiss Professor for AI in Neural Systems Imaging*

*Laboratory for Mental Health Mapping (MHM-Lab)*

*Friedrich Schiller University Jena, Jena, Germany*

*dr.thomas.wolfers@gmail.com*

[#] These authors contributed equally as shared last authors.

[*] Data used in preparation of this article were obtained from the Alzheimer's Disease Neuroimaging Initiative (ADNI) database (adni.loni.usc.edu). As such, the investigators within the ADNI contributed to the design and implementation of ADNI and/or provided data but did not participate in analysis or writing of this report. A complete listing of ADNI investigators can be found at: http://adni.loni.usc.edu/wp-content/uploads/how_to_apply/ADNI_Acknowledgement_List.pdf

[**] Data used in the preparation of this article was obtained from the Australian Imaging Biomarkers and Lifestyle flagship study of ageing (AIBL) funded by the Commonwealth Scientific and Industrial Research Organisation (CSIRO) which was made available at the ADNI database (www.loni.usc.edu/ADNI). The AIBL researchers contributed data but did not participate in analysis or writing of this report. AIBL researchers are listed at www.aibl.csiro.au.

## Abstract

Normative models chart how brain measures vary across the lifespan, expressing each individual as a deviation from a healthy reference. However, most regional models of cortical thickness disregard dependencies between cortical and subcortical anatomy, and transferring them to a new site typically requires retraining or a recalibration of location and scale alone. We condition lifespan norms of cortical thickness on subcortical volume, such that each deviation indicates whether an individual's cortex lies within the healthy range given their subcortical anatomy, and make them transferable. In 21,351 healthy individuals from 216 sites, conditioning on ten subcortical structures improved out-of-sample prediction on average, most for the amygdala and hippocampus. Refitting only the intercepts of all four distributional parameters on about 20 healthy individuals, as implemented in our open-source package NormGAMLSS, calibrated most of six unseen sites without retraining, for base and conditional models. In the Alzheimer's Disease Neuroimaging Initiative, global deviation and structural dissociation (the reduction in mean absolute deviation under conditioning) predicted conversion to mild cognitive impairment and dementia beyond the Clinical Dementia Rating and dissociation added beyond global deviation for dementia, even at similar clinically recorded impairment. Conditional norms thus provide relevant, individual-level information.

## Introduction

Brain charts have recently emerged as lifespan references for brain morphology. They build on normative modelling, which places an individual within the distribution of a reference population and turns a brain measure into an individual deviation score[1,2]. Such charts are now derived from large multi-site samples[3–5], and their deviation scores reveal heterogeneity that case-control averages conceal[6–8]. Generalized additive models for location, scale and shape (GAMLSS) capture how the centre, spread and shape of these measures change with age[9–11]. However, conventional brain charts typically model each cortical region in isolation, as a function of demographic and technical covariates such as age, sex and site and, at most, global measures such as intracranial volume[12–14]. The extent to which an individual's subcortical anatomy informs their cortical deviations therefore remains largely unexplored.

Subcortical structures are natural candidates for conditioning because cortex and subcortex are not independent: cortico–basal ganglia–thalamic loops connect them[15,16], the hippocampus connects directly with parahippocampal, perirhinal and temporal-pole cortex[17], and cortical and subcortical structures covary in morphology across individuals and during development[18,19]. A normative model conditioned on subcortical volume therefore addresses whether, given an individual's subcortical anatomy, cortical thickness remains within the expected healthy range. Alzheimer's disease illustrates the relevance of this question. In its typical form, neurofibrillary pathology begins in the medial temporal lobe and later reaches the neocortex[20], and models of its progression place medial temporal atrophy early[21]. Amygdala atrophy is comparable to hippocampal atrophy in the earliest clinical stages[22], and cortical thinning tracks symptom severity[23]. This sequence is not universal, however: the relationship between medial temporal and cortical atrophy varies across individuals, and this variation is associated with distinct clinical courses[24–27]. Normative deviation scores have been linked to progression from mild cognitive impairment to Alzheimer's disease[8,28]. Conditional models make it possible to test whether the component of an individual's cortical deviation that is accounted for by subcortical volume carries prognostic information beyond cognitive assessment.

Applying these models across sites poses a practical challenge. Scanner, acquisition protocol and field strength shift morphometric measures between sites[29–31]. Normative models often account for this heterogeneity by modelling site effects explicitly[3,4,32], in which case a site absent from training cannot be scored directly: the model must be refitted with that site included, or the site scored as though it resembled the training sites, which biases its centiles[11,12]. Feature-level harmonization offers another route, but standard ComBat estimates site effects jointly across the sites available at fit time, so a new site changes the harmonized features on which a model was trained[33–35]. Newer variants accommodate unseen sites but still adjust the features against a reference[36,37]. Alternatively, a fitted model can be recalibrated at the new site from a local sample, by re-estimating its site offsets while the rest of the model stays fixed[3,4,11] or, in hierarchical Bayesian models, by renewed Markov chain Monte Carlo sampling at the new site with the fitted model's posteriors as informative priors[35,38]. Reported sample requirements range from about 25 to more than 100 individuals, depending on the model and criterion[3,13,31,39]. However, these recalibrations typically adjust the location and scale of the estimated distribution while leaving its shape fixed[4,11,32], do not include site-level factors such as field strength in the adjustment, and have not been evaluated for models conditioned on other brain measures.

Here, we develop normative models of cortical thickness that are explicitly conditioned on subcortical volume and quantify the resulting predictive gain. Using a distributional modelling framework fitted to a lifespan cohort of 21,351 healthy individuals from 216 sites, we estimate a base model of regional cortical thickness alongside ten conditional models, each adding the volume of a single subcortical structure, and evaluate them on held-out individuals. All analyses are performed with NormGAMLSS, an open-source Python package released with this work that provides a normative-modelling interface to GAMLSS and implements the site-transfer procedure developed here. Conditioning improves out-of-sample prediction of cortical thickness on average for all ten structures, with the largest and most widespread gains for the amygdala and hippocampus. Applied at a site absent from training, these models require no retraining and do not inherit a proxy site's bias: refitting the intercepts of all four distributional parameters, optionally with site-level covariates such as field strength, on a small, age-stratified calibration sample removes this bias, and we establish the required sample size. Applied to an independent longitudinal cohort from the Alzheimer's Disease Neuroimaging Initiative (ADNI), the whole-brain z-score and structural dissociation, defined as the reduction in whole-brain mean absolute cortical deviation when the normative model is conditioned on subcortical volume, predict conversion to mild cognitive impairment (MCI) and to Alzheimer's disease dementia beyond cognitive testing, and for conversion to dementia dissociation adds beyond global atrophy. Among individuals with similarly mild clinical impairment, the imaging measures also distinguish those who later convert.

## Results

### Lifespan normative models capture age and sex effects on regional cortical thickness

We modelled cortical thickness across 68 Desikan–Killiany regions within a GAMLSS framework using a sinh-arcsinh (SHASH) outcome distribution, fitted separately for each region (Fig. 1a). The base model included age, sex, site, Euler number and intracranial volume. Ten conditional models were fitted, one for each subcortical structure returned by the segmentation - the thalamus, caudate, putamen, pallidum, hippocampus, amygdala, accumbens area, ventral diencephalon, cerebellar white matter and cerebellar cortex - each adding that structure's volume to the location parameter. A lifespan cohort of 21,351 healthy individuals across 216 sites was assembled for training, with evaluation on held-out healthy individuals from a test cohort that also contained a patient subgroup (Table 1; Fig. 1b; see Methods).

Among all 68 regions, the left transverse temporal gyrus and right parahippocampal gyrus showed the largest sex differences in estimated thickness (0.050 mm and 0.046 mm respectively, or approximately 0.29 and 0.25 of the fitted scale parameter σ, averaged across the training sample; Wald $|t| = 16.4$ and 14.4, both $q < 0.001$ after Benjamini–Hochberg correction across the 68 regions), and were clearly separated from the remaining regions (Fig. 1c; Supplementary Note 1). Because the age spline is shared between sexes, this offset is constant by construction: female thickness exceeded male thickness by the same margin at every age in both regions. In a sensitivity analysis, sex-specific age splines did not improve BIC or the out-of-sample log-score, although AIC favoured them in most regions, and under them the sex difference varied with age by a median of only 0.03 mm (Supplementary Note 8; Supplementary Table 4). In the conditional models, sex is in addition encoded indirectly through the conditioning volume, which itself differs between the sexes. The scale parameter also carries a sex effect through a log link, so the two sexes' centile bands stand in constant proportion to one another, narrowing through development to a minimum around the fourth decade and widening again in later life. These fits establish the reference behaviour of the base model against which every conditional model is subsequently compared.

***[Figure 1]***

**Conditioning on subcortical volume improves out-of-sample prediction of cortical thickness**

We next asked whether subcortical volume carries information about cortical thickness beyond the base model's covariates. For each of the 68 cortical regions, model fit was quantified by the out-of-sample log-score on held-out healthy individuals (see Methods).

Conditioning on subcortical volume improved out-of-sample log-predictive density on average for all ten structures, with the two medial temporal lobe structures showing the largest gains: the amygdala (mean ΔLog-score 0.0033; 90% of regions improved) and hippocampus (0.0032; 82%), followed by cerebellar cortex (0.0024; 76%), caudate (0.0016; 87%), accumbens area (0.0012; 79%), thalamus (0.0012; 78%), ventral diencephalon (0.00065; 62%), putamen (0.00048; 70%), pallidum (0.00026; 59%), and cerebellar white matter (0.00020; 49%) (Fig. 2a). Paired bootstrap 95% confidence intervals of the mean gain excluded zero for seven of the ten structures, all but putamen, pallidum and cerebellar white matter; for the amygdala, the gain corresponds to a 0.65% reduction in the residual variance of cortical thickness on average, and up to 3.9% in the most improved region (7.2% for the hippocampus) (Supplementary Note 8; Supplementary Table 5).

Across all 679 structure-by-region combinations, ΔLog-score was positive in 497 cases (73.2%), with a mean of 0.0014 and a median of 0.00056 (Fig. 2a,b). The largest single per-region improvement, 0.032, occurred for the hippocampus-conditional model of the right parahippocampal gyrus, an anatomically adjacent pairing. A complexity-penalized model-selection criterion computed independently on the training data placed the same three structures, the amygdala, hippocampus and cerebellar cortex, at the top, in a slightly different order (Supplementary Fig. 1). Stratified by age, the amygdala and hippocampus gains were similar below 20 years and at 20–60 years (amygdala 0.0031 and 0.0030; hippocampus 0.0030 and 0.0034), and in individuals under 20 both exceeded those of the caudate and putamen; the 69 individuals over 60 years were too few for a stable estimate (Supplementary Table 6). The hippocampus gain was the same in the better- and worse-quality halves of the scans, whereas the amygdala gain was concentrated in the worse-quality half (Supplementary Table 7). Subcortical volume also affected the spread of cortical thickness: added to the scale parameter of the amygdala and hippocampus models, it improved AIC in all 68 regions and BIC in 66 and 67, with a larger volume associated with a narrower distribution and a further held-out gain of about 0.002 (Supplementary Table 8).

These results establish, at the group level, that subcortical volume, particularly amygdala and hippocampal volume, carries information about cortical thickness beyond what age, sex, site, Euler number, and intracranial volume already explain. Because these structures are themselves reduced in neurodegenerative disease, the amygdala and hippocampus from the earliest clinical stages of Alzheimer's disease[22], conditioning on subcortical volume would be expected to absorb part of the cortical thinning already detectable in patients. Three further analyses followed from this. A region-wise comparison showed that the improvement was distributed across the cortex rather than driven by a few regions (Supplementary Fig. 2). Applying the models to an independent external cohort of healthy, MCI and dementia individuals showed that conditioning removed cortical regions from the set showing significant disease-related thinning and never added one, with the amygdala and hippocampus removing the most regions and largely the same ones (Supplementary Note 3; Supplementary Fig. 3; Supplementary Table 1). Quantifying how much of this thinning subcortical volume accounted for distinguished the two disease stages: nearly all of it in dementia, but only part of it in MCI (Supplementary Fig. 4).

***[Figure 2]***

**A small age-stratified calibration sample removes site bias without retraining**

A normative model that includes site as a covariate cannot be applied directly at a site absent from training: the site must either be added and the model retrained, or mapped onto a proxy site whose bias it then inherits. Feature-level harmonization offers a third route, but adjusts the data rather than the fitted model (see Methods). To test whether a small calibration sample can remove that bias instead, we evaluated site transfer at six held-out sites using the healthy-labelled subset of an independent external cohort (Fig. 3a). At such a site, differences in scanner, acquisition protocol and local population shift the whole distribution of thickness away from the model's estimate, so individuals are placed at systematically biased centiles even when the model's age and sex effects are themselves correct. Transfer corrects this by drawing a small, age-stratified sample of healthy individuals from the new site and refitting an intercept-only adjustment to the model's four distributional parameters, leaving everything else the model has learned intact, which returns the curve to the diagonal. For the base-versus-conditional comparisons in Fig. 3b,c, the hippocampus-conditional model served as the representative conditional model (Supplementary Note 4).

Site-transfer calibration improved performance at every held-out site, most steeply over the first 20 or so transfer individuals: at 20 individuals improvement was already positive at five of the six sites (+1.0% to +12.3%), compared with +15.7% (NYU-2), +10.0% (Site-34), +5.8% (SWU-4), +5.0% (OASIS3), +4.0% (Site-BRI), and +1.7% (NKIRS) at 100 individuals (Fig.

3d). A fivefold increase in calibration data beyond 20 individuals added a further 1.5 to 3.3 percentage points at each site. Mean squared error ranked the six sites in the same order and saturated over a similar range, and the amygdala- and hippocampus-conditional models showed the same qualitative pattern at closely comparable magnitudes (Supplementary Fig. 5). At the smallest samples tested the refit was unstable and could perform worse than no transfer at all, as expected when a distributional refit is estimated from very few observations (Supplementary Note 4). Across 50 repeated age-stratified draws per site, the gain at 20 individuals was positive in 84% of draws for the base model and 87% for the hippocampus-conditional model, and in every draw at four of the six sites, reaching a median of 57% and 58% of the gain obtained with 100 individuals in the same draw (Supplementary Note 8; Supplementary Table 11).

With a 20-individual transfer sample, the median percentage improvement in log-score across the 68 regions was +3.6% for the base model and +3.4% for the hippocampus-conditional model (means +4.7% and +4.6%), indicating the procedure generalizes to a conditional model at a comparable magnitude (Fig. 3b,c). At the same calibration sizes, refitting the location intercept only, the location and scale intercepts, or all four intercepts gave median log-score gains of 6.0%, 5.3% and 3.6% at 20 individuals and 7.3%, 7.7% and 7.8% at 100 (Supplementary Table 10). With few calibration individuals, refitting the location intercept alone therefore helped most, and the full refit matched or exceeded it from about 50 individuals, consistent with the location-dominated site effects described next.

The z-score distribution shows what the calibration is correcting, and the two parameters behave differently (Fig. 3e). Before any transfer, every site's mean z-score was negative, ranging from −0.19 (NKIRS) to −0.89 (NYU-2). Thus, the model systematically overestimated thickness at uncalibrated sites. The standard deviation, by contrast, was already close to its target of 1 at every site (0.896 to 1.031). Calibration therefore primarily needed to correct location, whereas scale was already largely correct. With 20 transfer individuals the mean had moved to within 0.34 of zero at every site, and by 100 individuals only to within 0.19, while the standard deviation returned to between 0.945 and 1.100. Both parameters therefore settle over the same range as the log-score, and both show the same instability at the smallest samples (Supplementary Note 4).

Together these results support treating a small, age-stratified sample of around 20 individuals from a new site as sufficient to bring most sites into calibration, for both the base and conditional models, at a fraction of the data a retrained model would need. Beyond this point, additional calibration data yielded smaller further gains. The main caveat is that the magnitude of the gain varied considerably between sites (Supplementary Note 4). The same refit procedure also addresses scanner field-strength bias, which is substantial if left uncorrected; correcting it is a prerequisite for the clinical application in Fig. 4, since ADNI is a new site with mixed field strengths. Placing field strength on all four distributional parameters reduced this bias more than placing it on location alone, at both calibration budgets tested (Supplementary Note 5; Supplementary Fig. 6; Supplementary Table 2).

***[Figure 3]***

**Subcortical–cortical structural dissociation predicts conversion beyond cognitive testing and global atrophy**

We finally evaluated whether structural dissociation, the reduction in whole-brain mean absolute cortical deviation when the normative model is conditioned on subcortical volume, carries prognostic information in ADNI (Fig. 4a), a cohort held out entirely from normative model estimation and brought into calibration using the transfer procedure above. Survival and classification models were fitted within ADNI and evaluated out-of-fold (see Methods).

Across all ten structures and both transitions, adding the whole-brain z-score and structural dissociation to the Clinical Dementia Rating Sum of Boxes (CDR-SB) improved out-of-fold concordance (Fig. 4a): every one of the twenty structure-by-transition combinations gained significantly over CDR-SB alone, from 0.602 to as high as 0.678 for CN→MCI and from 0.697 to 0.740 for MCI→AD. Because the whole-brain z-score contributes on its own whichever structure accompanies it, this comparison does not discriminate between structures. Accordingly, no structure is highlighted in Fig. 4a, and the structures carried forward were selected by the stricter test described in Supplementary Note 6.

Hippocampal and amygdala dissociation identified largely the same individuals, consistent with the two structures capturing a shared signal. We therefore present the hippocampus in the main figures and treat the amygdala as a robustness check (Supplementary Note 6). The similarity held in the Kaplan–Meier curves: the hippocampus model separated predicted risk groups sharply (Fig. 4c; 261 individuals per group; 46 vs. 134 events; log-rank $P = 3.6 \times 10^{-23}$), and the amygdala model gave an almost identical separation (same group sizes and event counts; log-rank $P = 3.3 \times 10^{-22}$; Supplementary Fig. 7a). The hippocampus model was also reasonably calibrated and, in decision-curve analysis, gave a higher net benefit than CDR-SB alone at risk thresholds of 15–55% for 3-year and 10–45% for 2-year risk (Supplementary Note 8; Supplementary Table 13). Applied to the earlier CN→MCI transition using ventral diencephalon, the full pipeline found a markedly weaker effect (Supplementary Fig. 8). Selecting the structure inside the cross-validation folds instead of in the whole cohort confirmed the MCI→AD result (amygdala or hippocampus chosen in every fold; out-of-fold gain in concordance 0.014, 95% CI 0.002–0.028), whereas the CN→MCI gain did not persist (Supplementary Note 8; Supplementary Table 12).

The landmark analysis showed that this pattern persisted, albeit attenuated, as follow-up accumulated (Fig. 4b). At the earliest landmark (573 individuals, 180 converters), adding hippocampal dissociation improved discrimination by 0.055 (95% CI 0.022–0.088), and adding amygdala dissociation by 0.047 (95% CI 0.015–0.082). At the second landmark (257 individuals, 95 converters), the improvement was 0.041 for hippocampus (95% CI 0.003–0.086) and 0.033 for amygdala (95% CI −0.004 to 0.078). By the two latest landmarks (180 and 124 individuals), point estimates remained positive (0.031-0.041 for both structures) but confidence intervals crossed zero, consistent with reduced power at smaller samples rather than a reversal. Visit-by-visit trajectories, which locate this signal across follow-up and trajectory groups, are shown in Supplementary Fig. 9.

Within the MCI→AD cohort, converters showed higher whole-brain z-scores and higher hippocampal dissociation than stable individuals in the two lowest bins with testable sample size (Fig. 4d,e; $n = 250$ and $n = 115$), with the difference no longer significant in the two higher bins ($n = 149$ and $n = 64$). The lowest bin, CDR-SB = 0, contained two individuals and was not tested.

Because individuals within a bin have the same or closely similar baseline cognitive scores, this separation largely reflects information not captured by CDR-SB: among individuals who appear similar on clinical assessment, the imaging measures still identified those who subsequently converted. The separation was present in the two mildest bins, where the cognitive score is least informative, and disappeared once impairment was sufficiently marked for CDR-SB to separate the groups on its own. The attenuation is not explained by sample size, and the population used here is defined differently from the Cox survival cohort (Supplementary Note 6). Following these two bins visit by visit showed that the separation was already present at the earliest available visit rather than emerging as conversion approaches (Supplementary Fig. 10).

An alternative interpretation is that dissociation is not an early signal but merely a marker of how close an individual already was to converting when baseline data were collected. At baseline, hippocampal dissociation did not differ detectably between individuals who converted within 1.5 years and those who converted later, but was higher in converters than in stable individuals (Supplementary Note 6), consistent with a distinction by eventual outcome rather than by proximity to the event.

***[Figure 4]***

## Discussion

We built normative models of regional cortical thickness conditioned on the volume of a single subcortical structure, transferred them to unseen sites from a small calibration sample, and used the difference between base and conditional deviations to predict clinical progression. Conditioning improved out-of-sample prediction on average for all ten subcortical structures, most for the amygdala and hippocampus. Refitting the intercepts of all four distributional parameters on about 20 age-stratified individuals brought most of six held-out sites into calibration for base and conditional models alike, and the same procedure with field strength as a covariate reduced scanner bias in ADNI. There, the whole-brain z-score and structural dissociation added to CDR-SB in predicting conversion to MCI and to dementia; for conversion to dementia, dissociation also added beyond the whole-brain z-score and separated converters from stable individuals among those with the mildest clinical impairment. Conditional norms can therefore be moved across sites at low cost, and the component of an individual's cortical deviation accounted for by subcortical volume carries clinically relevant information.

### Conditioning cortical norms on subcortical volume

The gain from conditioning was largest for the hippocampus and amygdala, and the largest single improvement paired the hippocampus with the anatomically adjacent right parahippocampal gyrus. The human hippocampus connects directly with parahippocampal, perirhinal and temporal-pole cortex[17], and hippocampal volume covaries most strongly with medial temporal and other memory-system regions across individuals[18]. Part of what the conditional models capture is therefore likely to reflect regional anatomy rather than overall head size, which intracranial volume already absorbs in the base model. Cerebellar cortex ranked third, in keeping with the extensive mapping of cerebellar territories onto cerebral association cortex[40], although that evidence is functional rather than morphometric. Although the average gains were small, they were positive in 73% of structure–region combinations, as expected when one covariate refines a model that already contains age, sex, site and intracranial volume. Within this lifespan sample, the medial temporal advantage did not depend on age: the amygdala and hippocampus gains were similar below 20 years and at 20–60 years (Supplementary Table 6). Which structure is most informative may still differ between disorders, so comparing structures within the population of interest is a better guide than a fixed choice. Subcortical volume also narrowed the estimated spread of cortical thickness when added to the scale parameter (Supplementary Table 8), so conditional models that carry it on both location and scale are a natural extension.

Conditioning also changes what a deviation means. Adjusting for a variable that a disease itself alters can absorb part of the disease effect[41], and this is what we observed: in dementia, the cortical shift produced by conditioning was almost entirely what the healthy relationship between subcortical volume and cortex implies, whereas in MCI a modest part of it was not (Supplementary Fig. 4). A conditional model is therefore not a more sensitive detector of cortical thinning. Rather, it asks whether cortex is thinner than expected given the same individual's subcortical volumes. In MCI, the regions accounted for by amygdala or hippocampus conditioning were the insula, the left temporal pole and the right parahippocampal and lingual gyri, and the temporal pole is among the regions in which thinning has been reported in prodromal Alzheimer's disease[23,42].

This differs from two existing ways of going beyond single-measure norms. Regional models that include global measures such as intracranial volume or mean thickness adjust for overall size[12], as our base model does; conditioning adds a specific structure on top of it. Multivariate and deep normative models learn the joint structure of many brain measures but condition on or adjust for demographic covariates[14,43,44], whereas conditioning one region on one named structure keeps each deviation a centile of an explicit conditional distribution. The software landscape is similarly complementary (Supplementary Note 9; Supplementary Table 15), and the tools differ most in what a new site must supply before its individuals can be scored. PCNtoolkit offers Bayesian linear and hierarchical Bayesian regression with transfer, extension and merging of fitted models[35,45,46]; the Lifespan Brain Chart and CentileBrain resources provide pretrained reference models, adapted to a new study by re-estimating its offsets or applied after harmonizing the user's data, respectively[3,47]; and Brain MoNoCle offers a no-code interface with a site calibration of the mean and, given at least 30 healthy individuals, of the standard deviation, keeping skewness and kurtosis from the trained model[32]. A clinical site, however, typically has few local healthy controls, may combine scanners or field strengths, as ADNI does, and flags individuals by their position in the tails of the distribution, which depends on its scale and shape as well as on its location. NormGAMLSS provides a Python interface to the full range of GAMLSS distributions and per-parameter formulas[9,48], which conditional models require, and a site transfer suited to this setting: it refits any subset of the four distributional parameters, optionally with site-level covariates such as field strength (Supplementary Note 5), and about 20 healthy individuals brought most held-out sites into calibration (Supplementary Notes 4 and 8.8; Code availability).

**Transferring normative models to unseen sites**

Approximately 20 age-stratified individuals yielded a consistent gain at an unseen site, positive in most repeated calibration draws and about half to three-quarters of the gain obtained with five times as many (Supplementary Table 11). Earlier work on adapting normative models has reported requirements from about 25 to more than 100 individuals[3,13,31,39]. These figures are not directly comparable: the studies differ in the model adapted, in what is adjusted, from random-effect offsets to mean shifts or informative priors, and in the criterion, from the stability of centile scores and the plateau of fit metrics to the validity of reference intervals. Two features of the present design may contribute to the lower figure. Calibration individuals were drawn across the site's age range, and the age distribution of an adaptation set has been reported to affect adapted models[39]. The refit also estimates only intercepts, with the trained model entering as an offset, so each distributional parameter costs a single coefficient. Site-specific skewness is hard to estimate from small samples[32]: our four-parameter refit was unstable at five individuals, and with up to about 30 individuals refitting the location intercept alone gave larger gains, whereas from about 50 individuals the full refit matched or exceeded it (Supplementary Table 10). For a covariate that acts on the whole distribution, in contrast, adding field strength to all four parameters reduced scanner bias in ADNI more than adding it to location alone (Supplementary Fig. 6). Before transfer, every site's mean z-score was negative while its standard deviation was already close to one, so in these data site effects acted mainly on location, although effects of site on scale are documented elsewhere[11].

Site transfer also differs in level from harmonization. Standard ComBat estimates site effects for each feature jointly across the sites present at fit time[33,35], so a model trained on harmonized features has to be refitted when a site is added; recent variants harmonize a new site against a fixed reference instead[36,37]. Site transfer leaves the features and the trained model unchanged and adjusts only the model's distributional parameters at the new site, and a benchmark against standard ComBat would therefore compare different objects. The size of the calibration sample matters for translation: centile scoring for small studies or single patients has been described as a goal for future work[3], and a requirement of about 20 local healthy individuals, without retraining, places it within reach of a single clinical site. This is feasible because the transfer re-estimates only a few site intercepts of an already well-estimated model, whereas estimating the outer centiles of a normative distribution from scratch requires far larger samples, of the order of thousands of individuals[49]. The procedure is also computationally inexpensive:

refitting all four parameters took under 7 s per region on a single processor core, so all 68 regional models of a site can be transferred in under ten minutes on one core. Inexpensive recalibrations elsewhere shift the mean and rescale the standard deviation[32,45], whereas adapting the full distribution in hierarchical Bayesian regression re-runs Markov chain Monte Carlo sampling at the new site[35,45], which its developers describe as computationally demanding, although less so with fewer batch effects[38]. Site transfer adjusts the full distribution at the cost of an inexpensive refit.

**Structural dissociation and conversion**

Structural dissociation measures how much of an individual's brain-wide cortical deviation a subcortical volume accounts for. For MCI→AD, hippocampal and amygdala dissociation each added to a baseline of CDR-SB and whole-brain z-score, and the two identified largely the same individuals, consistent with comparable atrophy of the two structures in early Alzheimer's disease[22] and with co-pathologies, such as limbic-predominant age-related TDP-43 encephalopathy, that affect both[50]. The measure lends itself to interpretation in terms of atrophy subtypes. The balance between hippocampal and cortical atrophy distinguishes limbic-predominant, typical and hippocampal-sparing presentations with different clinical courses[24–26], and in ADNI a hippocampal-to-cortical volume ratio was associated with the rate of clinical decline, mostly through cortical atrophy[27]. Dissociation carries this idea from group-defined ratios to an individual, model-based quantity referenced to a healthy lifespan population.

Normative deviation scores and brain-age gaps have been associated with progression from MCI[8,28,51]. We asked what such measures add to a strong clinical baseline, since CDR-SB is among the strongest single predictors of progression[52,53]. Within a severity bin, individuals have the same or closely similar baseline CDR-SB, yet converters still differed on the imaging measures in the two mildest bins that could be tested, whereas at higher scores the cognitive score separated the groups on its own. The bin analysis does not by itself isolate dissociation from global atrophy, because the whole-brain z-score separated converters in the same bins; that distinction rests on the likelihood-ratio test of dissociation added to CDR-SB and whole-brain z-score, its repetition with the structure selected inside the cross-validation folds, and the landmark comparison (Fig. 4b; Supplementary Note 6; Supplementary Tables 3 and 12). As with residual measures in general[54], a difference between two model-based summaries is expected to correlate with the measures it is derived from, which is why the survival models included the whole-brain z-score alongside dissociation. Baseline dissociation also did not differ detectably by time to conversion but did differ by eventual outcome (Supplementary Note 6), which does not support it simply marking proximity to conversion.

The CN→MCI transition was harder to characterize: the same pipeline applied with the ventral diencephalon gave a markedly weaker effect (Supplementary Fig. 8), with 34 events among 184 individuals, and the gain did not persist when the structure was selected inside the cross-validation folds (Supplementary Table 12). The small number of converters and the subtlety of structural change before symptom onset, when cortical thinning can precede hippocampal volume loss[55], both limit what a baseline structural measure can predict at this stage.

**Strengths and limitations**

The study combines a large lifespan training cohort processed with a single FreeSurfer version, matched comparisons of base and conditional models at the individual level, evaluation at sites and in cohorts absent from training, and out-of-fold evaluation of every clinical model. Several limitations should be noted. First, ADNI diagnoses are clinical rather than biomarker-defined[56,57], so conversion denotes a syndromic transition. The cognitively normal individuals used to calibrate ADNI were likewise not selected on amyloid status and may include some with preclinical Alzheimer's disease, which would move the calibrated norm towards early atrophy and, if anything, attenuate deviations. Second, models fitted to cross-sectional data were applied to longitudinal visits; they locate an individual relative to a population rather than model individual change[58]. Third, large reference datasets over-represent European and North American populations[3], and the representativeness of the present cohort was not assessed. Fourth, Wald standard errors for parametric terms in GAMLSS models with smooth terms are approximate[48], so the sex-difference tests are descriptive. Fifth, cortical thickness and subcortical volume were derived from the same scan and FreeSurfer run, so part of the conditioning gain may reflect shared measurement variance from scanner, motion or segmentation quality, which the Euler number captures only in part; the hippocampus gain did not depend on scan quality, but the amygdala gain was concentrated in the lower-quality half of the scans (Supplementary Table 7). Finally, conditional models are predictive and imply no causal direction between subcortical and cortical structure.

## Conclusion

Cortical thickness is not independent of subcortical anatomy, and conditional norms make that dependence measurable in each individual. Instead of asking whether an individual's cortex is unusual for their age and sex, they ask whether it is unusual given that individual's own subcortical anatomy. In Alzheimer's disease, this distinction separates the disease stages. In dementia, cortical thinning followed almost entirely from subcortical atrophy; in mild cognitive impairment, part of it did not. The resulting dissociation identified who would progress to dementia beyond clinical severity and global atrophy, already at the earliest visits. Because these norms transfer to new sites without retraining, any clinic or cohort can apply them.

## Online Methods

### Cohorts

Demographic characteristics of all cohorts are given in Table 1. Models were trained on a healthy developmental-to-ageing cohort ($n$ = 21,351; 216 sites; ages 3–100). Evaluation used a held-out healthy sample ($n$ = 2,176; 79 sites; ages 5–79) drawn from a separate test cohort. That same test cohort also contained a patient subgroup ($n$ = 3,684; 64 sites; ages 5–94) spanning seven diagnostic categories: attention-deficit/hyperactivity disorder ($n$ = 1,568), autism ($n$ = 1,074), major depressive disorder ($n$ = 251), Alzheimer's disease ($n$ = 205), schizophrenia ($n$ = 199), learning disorders ($n$ = 197), and generalized anxiety disorder ($n$ = 190). The patient subgroup was excluded from every evaluation. It was markedly younger than its 5–94-year range suggests (median 12.5 years, IQR 9.2–23.0), reflecting the predominance of attention-deficit/hyperactivity disorder and autism, which made up 72% of the subgroup and had median ages of 10.2 and 13.9 years. The training and test cohorts were assembled from 117 existing datasets, listed with their numbers of individuals in Supplementary Data 1. HCP Young Adult, which includes twins and siblings, contributed only to the training set. Data from the Cambridge Centre for Ageing and Neuroscience were obtained from the CamCAN repository (https://cam-can.mrc-cbu.cam.ac.uk/dataset/), based on the CamCAN study[59].

All T1-weighted scans, in every cohort, were processed with FreeSurfer 7.4.1[60]. Cortical thickness[61,62] was taken from the 68 regions of the Desikan–Killiany atlas[63], and subcortical volumes from FreeSurfer's automated segmentation[64], averaged across the left and right hemispheres, so that each structure provides one conditioning variable for cortical regions in both hemispheres; conditioning on the ipsilateral volume instead gave comparable results (hippocampus: no significant difference; amygdala: a small reduction, 0.0028 vs 0.0033;

Supplementary Table 9). Intracranial volume was FreeSurfer's estimated total intracranial volume[65], and the Euler number was summed over the two hemispheres' reconstructed surfaces.

Data used in the preparation of this article were obtained from the Alzheimer's Disease Neuroimaging Initiative (ADNI) database (adni.loni.usc.edu). The ADNI was launched in 2003 as a public-private partnership, led by Principal Investigator Michael W. Weiner, MD. The primary goal of ADNI has been to test whether serial magnetic resonance imaging, positron emission tomography, other biological markers, and clinical and neuropsychological assessment can be combined to measure the progression of mild cognitive impairment and early Alzheimer's disease. We used individuals enrolled in ADNI-1, ADNI-GO and ADNI-2[57,66], scanned at 1.5 T or 3 T[67,68] (2,830 and 2,245 visits, respectively); imaging data were downloaded on 4 March 2024 and clinical data on 14 November 2024. The ADNI cohort used for the clinical analyses comprised 5,075 visits from 1,229 individuals; where a visit had more than one scan (604 visits, all repeat scans at the same field strength), one scan per visit was used. Each individual's trajectory was defined from the diagnoses recorded at the first and last visit carrying usable z-scores, which were available for every individual: 420 remained MCI, 341 remained cognitively normal, 230 remained demented, 194 converted from MCI to dementia, 37 from cognitively normal to MCI, and 7 from cognitively normal to dementia. Median follow-up was 2.0 years (interquartile range 1.0–4.0, maximum 10.8); mean age at first visit was 73.7 years (SD 7.1, range 54.4–89.6) and median baseline Clinical Dementia Rating Sum of Boxes (CDR-SB)[69,70] was 1.0.

**Normative modelling framework**

Cortical thickness across 68 Desikan–Killiany regions was modelled with a GAMLSS framework[9] using the sinh–arcsinh (SHASH) outcome distribution as implemented in the R gamlss.dist package[71], fitted separately for each region. In the base model, the location parameter was modelled as a penalized B-spline of age[72] together with a main effect of sex, a random intercept for site, Euler number (a surrogate for scan and reconstruction quality[73]), and intracranial volume; the scale parameter was modelled as a penalized B-spline of age plus a main effect of sex; and the skewness and kurtosis parameters were each estimated as a single constant per region. The location parameter used an identity link, and the scale, skewness, and kurtosis parameters each used a log link. We refer to this model as the base model. The age spline of the location parameter is shared between the sexes; sex-specific splines, fitted with `pvc(age, by = sex)` because the penalized B-spline term in gamlss takes no `by` argument, were examined in a sensitivity analysis (Supplementary Note 8).

Formally, for individual $i$ scanned at site $s(i)$, the base model for regional cortical thickness $y_i$ is

$$y_i \sim \mathrm{SHASH}(\mu_i, \sigma_i, \nu, \tau) \tag{1}$$

$$\mu_i = f_\mu(\mathrm{age}_i) + \beta_1\,\mathrm{sex}_i + \beta_2\,\mathrm{EN}_i + \beta_3\,\mathrm{ICV}_i + b_{s(i)}, \quad b_s \sim N(0, \sigma_b^2) \tag{2}$$

$$\log \sigma_i = f_\sigma(\mathrm{age}_i) + \gamma_1\,\mathrm{sex}_i \tag{3}$$

$$\log \nu = \nu_0, \quad \log \tau = \tau_0 \tag{4}$$

where $\mu$, $\sigma$, $\nu$ and $\tau$ are the location, scale, skewness and kurtosis parameters of the SHASH distribution, $f_\mu$ and $f_\sigma$ are penalized B-splines of age, $b_s$ is a random intercept for site, EN is the Euler number and ICV is intracranial volume. Models were fitted by penalized maximum likelihood with the RS algorithm[9], in R 4.1.2[74] with gamlss 5.4-22[48], called from Python through NormGAMLSS and rpy2 3.5.17[75]. An individual's deviation (z-score) is the standard normal quantile of the fitted cumulative distribution function at the observed value, a quantile residual[76], with the probability bounded at $10^{-6}$ from 0 and 1; it is computed in the same way for base, conditional and transferred models. Sex differences in estimated thickness were tested with a Wald test on the sex coefficient of the location parameter[48], with Benjamini–Hochberg correction across the 68 regions[77].

Conditional models extended this specification with one additional covariate on the location parameter: the volume of one of ten subcortical structures (the thalamus, caudate, putamen, pallidum, hippocampus, amygdala, accumbens area, ventral diencephalon, cerebellar white matter, and cerebellar cortex). Every other term, and the models for $\sigma$, $\nu$ and $\tau$, were left unchanged, so that for structure $k$ with volume $V_k$,

$$\mu_i = f_\mu(\text{age}_i) + \beta_1\, \text{sex}_i + \beta_2\, \text{EN}_i + \beta_3\, \text{ICV}_i + \beta_4\, V_{k,i} + b_{s(i)} \qquad (5)$$

This resulted in one base model and ten conditional models. Variable rescaling applied for numerical stability, and the construction of the normative-fit panel in Fig. 1c, are given in Supplementary Note 1.

**Model comparison**

Models were fitted on the healthy training cohort of 21,351 individuals and compared on the healthy subset of the held-out test cohort of 2,176 individuals, with that cohort's patient subgroup excluded from every evaluation. For each of the 68 cortical regions, the base model was compared against ten conditional models, each adding one subcortical structure's volume as an additional covariate. Base and conditional models were fitted on identical training data and evaluated on identical held-out individuals, so every comparison is matched at the individual level.

Model fit was quantified by the log-score, a strictly proper scoring rule[78]: the log of the probability density that a fitted model assigns to an individual's observed thickness, so a model scores higher the more density it places at the value actually seen. Because this density is evaluated from all four SHASH parameters, the log-score is sensitive to the width and shape of the predicted distribution and not only to its centre. Log-scores were computed per region as the mean across held-out individuals. ΔLog-score is the conditional model's log-score minus the base model's; positive values indicate a better-calibrated, sharper predictive distribution. Uncertainty in ΔLog-score was quantified by a paired bootstrap over held-out individuals, and its size was also expressed as the reduction in residual variance; these and further sensitivity analyses (age bands, scan quality, volume on the scale parameter, ipsilateral volume) are described in Supplementary Note 8. Handling of extreme observations, and the one structure–region fit excluded for non-convergence, are given in Supplementary Note 2.

**Site transfer**

Site transfer addresses the problem of applying a trained model at an out-of-distribution site, one absent from training, where the usual alternatives are to retrain the model with that site included or to assign the site to a proxy and inherit its bias[11,12].

ComBat harmonization[33] offers a third route, but estimates each feature's site effects jointly across the sites present at fit time, so accommodating a new site alters the training features and in turn the model already fitted to them[35]. In a direct comparison in a cohort aged 6–40 years, modelling site within a hierarchical normative model gave better predictive performance than ComBat harmonization[79]. Our proposed site transfer leaves both the features and the trained model intact, refitting only its distributional parameters. Fig. 3a illustrates this with observed-versus-expected centile curves, which plot the proportion of individuals actually falling below each nominal centile against that centile, a check of probabilistic calibration[80]. A correctly calibrated model traces the diagonal, because 10% of individuals should sit below the 10th centile, 25% below the 25th, and so on; an uncalibrated model bows away from it, and the direction of the bow says whether the model over- or underestimates at that site. Fig. 3a depicts that procedure schematically rather than plotting fitted data.

For a new site, a small subsample of that site's own healthy individuals was used to refit an intercept-only adjustment to each of the model's four distributional parameters, holding the trained model otherwise fixed: the location parameter adjusted additively, and the scale, skewness, and kurtosis parameters each multiplicatively, consistent with their log link. The adjusted model was then evaluated on the remaining individuals at that site, who were not used in the refit. On one core of an Intel Xeon Gold 6230R processor, a refit took 4.8–6.5 s per region, including loading the trained model (two regions at OASIS3, with 20 and 100 calibration individuals). The refit is deliberately minimal: only the intercepts move, so the site adjustment cannot absorb or distort the age trajectories the model was trained to represent. This procedure builds on offset-based recalibration[11], generalizing it to all four distributional parameters, or any subset of them, and to site-level covariates; it is implemented in NormGAMLSS, and full details are given in Supplementary Note 7, with a step-by-step protocol in Supplementary Note 10.

Site-transfer performance was evaluated at six held-out sites not part of model training (OASIS3, NKIRS, SWU-4, NYU-2, Site-34, Site-BRI), using the healthy-labelled subset of an independent external cohort. Site-BRI, and the Royal Melbourne Hospital 3T site used in Supplementary Note 3, are from the Australian Imaging Biomarkers and Lifestyle flagship study of ageing (AIBL), obtained through the LONI Image and Data Archive; the imaging data were downloaded on 2 February 2024 and the diagnostic groups were taken from the image collection list of 4 March 2024. Data was collected by the AIBL study group. AIBL study methodology has been reported previously[81]. Refitting was repeated at a fixed sequence of transfer-sample sizes (5 to 100 individuals, at 5-individual increments up to 50 and 10-individual increments thereafter) plus a no-transfer baseline. Calibration individuals were drawn age-stratified, as in every other transfer analysis in this paper; the full set of transfer settings used across analyses is given in Supplementary Note 4. The site-inclusion threshold, the fixed calibration pool, the choice of the hippocampus-conditional model as the representative conditional model, and the panel-level derivations for Fig. 3b–e are also given there. The comparison with refitting fewer distributional parameters, and the variability of the gain across repeated calibration draws, are described in Supplementary Note 8.

**ADNI analyses: cohort construction**

Two transitions were analysed. The CN→MCI analysis asks who converts among individuals cognitively normal at baseline, so its cohort combines both outcomes of that starting point: the 341 who remained cognitively normal and the 37 who converted to MCI, 378 individuals in all. The MCI→AD analysis is built the same way from the 420 who remained MCI and the 194 who converted to dementia, 614 in all.

Site-transfer calibration for this cohort drew 170 individuals from a pool consisting of exactly the 341 stable cognitively normal individuals (Supplementary Note 4). Those 170 therefore sit entirely within the non-converting arm of the CN→MCI cohort and were excluded from it. Requiring a baseline CDR-SB and positive follow-up time removed a further 24 (10 and 14, respectively), leaving 184 individuals with 34 events. No MCI individual was ever eligible for calibration, so no calibration exclusion applies to the MCI→AD cohort, where the same two requirements reduced 614 candidates to 522 individuals with 180 events (34 without a baseline CDR-SB and 58 without positive follow-up time).

**ADNI analyses: dissociation and survival modelling**

For each individual and visit, a whole-brain z-score summary was computed as the mean absolute cortical z-score across all 68 regions from the base model; the same summary was computed from each conditional model, and dissociation was defined as the base summary minus the conditional summary. In the held-out healthy sample, dissociation was centred on zero (hippocampus: mean −0.0001, SD 0.026; amygdala: mean 0.0002, SD 0.028), and a signed version, computed from mean rather than mean absolute z-scores, gave the same picture (Supplementary Table 14). Baseline associations with time to conversion were assessed with Cox proportional-hazards models[82], evaluated by out-of-fold concordance[83] under stratified 5-fold cross-validation, comparing CDR-SB alone against CDR-SB plus whole-brain z-score, CDR-SB plus dissociation, and CDR-SB plus both, with nested comparisons by likelihood-ratio test. Kaplan–Meier[84] risk groups were defined by a median split of out-of-fold risk from the joint model and compared by log-rank test[85]. Calibration and clinical utility of the MCI→AD model were assessed by decile calibration and survival decision-curve analysis (Supplementary Note 8). Plotting conventions and the age and sex adjustment for the trajectory panels (Supplementary Figs. 8b, 9 and 10) are given in Supplementary Note 6.

**ADNI analyses: landmark analysis (Fig. 4b)**

The landmark analysis used out-of-fold logistic regression (ROC AUC[86], same cross-validation scheme) to predict eventual conversion from features re-derived at each of four progressively later landmarks. Following the landmark approach[87,88], each landmark used its own eligible cohort: individuals with usable data at every visit up to that landmark and, for converters, no conversion by then. This reports AUC, not concordance. Its reference model already incorporates the rate of cognitive change, and is specified in Supplementary Note 6.

**ADNI analyses: severity-bin comparison (Fig. 4d,e)**

Individuals were grouped by their baseline CDR-SB, with a fixed bin for a score of zero and the non-zero remainder divided at quantiles of the observed distribution. Within each bin, converters and stable individuals were compared on whole-brain z-score (Fig. 4d) and on hippocampal dissociation (Fig. 4e) by an independent two-sample t-test assuming equal variances, performed only where both groups contained at least five individuals. The rationale for stratifying this way is given in Supplementary Note 6.

**ADNI analyses: structure selection**

Two comparisons were made for every structure. The first, CDR-SB alone against CDR-SB plus whole-brain z-score and dissociation, is shown in Fig. 4a and was passed by every structure. The second, a likelihood-ratio test of dissociation added to CDR-SB and whole-brain z-score, selected the structures used in the individual-level analyses: the hippocampus and amygdala for MCI→AD (Fig. 4b,c,e) and the ventral diencephalon for CN→MCI (Supplementary Fig. 8). Both comparisons, with full statistics for all ten structures in both trajectories, are given in Supplementary Note 6 and Supplementary Table 3. To gauge the optimism of selecting in the same cohort, the second comparison was also repeated inside the training folds of the cross-validation (Supplementary Note 8).

**Software**

Normative models were fitted and transferred with NormGAMLSS (R 4.1.2 and gamlss 5.4-22 through rpy2 3.5.17). Survival analyses used lifelines 0.30.0[89], cross-validation, logistic regression and ROC AUC used scikit-learn 1.5.2[90], and statistical tests used SciPy 1.14.1[91]. Cortical maps were drawn with python-ggseg[92].

## Data availability

All data analysed in this study come from existing datasets; no new data were collected. Supplementary Data 1 lists every dataset with its role, numbers of individuals, access point, dataset DOI, licence and the citation or acknowledgement it requests. The training and test cohorts comprise 117 datasets, 98 of them openly available on OpenNeuro (https://openneuro.org) under the accession numbers given there, including the Transdiagnostic Connectome Project (ds005237). The others are available from their providers, under their data use agreements where these apply: the Philadelphia Neurodevelopmental Cohort (dbGaP phs000607.v3.p2)[93]; the Brain Genomics Superstruct Project (Harvard Dataverse, doi:10.7910/DVN/25833)[94,95]; the Human Connectome Project Young Adult[96] and Aging[97,98] studies; HCP Development (doi:10.15154/1520708)[98,99], HCP Early Psychosis (doi:10.15154/1522899)[100] and the Boston Adolescent Neuroimaging of Depression and Anxiety study (doi:10.15154/3tk5-pb47)[101], through the NIMH Data Archive; the Cambridge Centre for Ageing and Neuroscience[59]; IXI (https://brain-development.org/ixi-dataset/); the Healthy Brain Network[102]; the SRPBS multi-disorder database[103]; OASIS-4[104]; the Pediatric Imaging, Neurocognition and Genetics study; and ABIDE, ABIDE II, ADHD-200[105], COBRE, ACPI and the 1000 Functional Connectomes Project[106], through the International Neuroimaging Data-sharing Initiative (https://fcon_1000.projects.nitrc.org). The external sites came from OASIS-3[107], the NKI-Rockland Sample[108], the Consortium for Reliability and Reproducibility[109], the Parkinson's Progression Markers Initiative (https://www.ppmi-info.org; RRID:SCR_006431) and the Australian Imaging, Biomarker and Lifestyle study (https://aibl.org.au), and the clinical cohort from ADNI (https://adni.loni.usc.edu). The fitted normative models are not publicly distributed at present; researchers can instead have their own data scored against them on the authors' computing infrastructure, without the data or the models being shared. Individual-level data cannot be redistributed under the data use agreements and are available from the original providers. Source Data are provided with this paper.

## Code availability

NormGAMLSS, the open-source Python package used for all normative modelling and site transfer in this study, is available at https://github.com/NormGAMLSS/NormGAMLSS, with documentation at https://normgamlss.github.io/NormGAMLSS/. A step-by-step protocol for fitting, evaluating and transferring models with NormGAMLSS, with code, is given in Supplementary Note 10. The version used here is release v0.2.0. Software versions are given in Online Methods (Software).

## Acknowledgements

T.W. was supported by the German Research Foundation (DFG) through an Emmy Noether grant (513851350), by the Federal Ministry of Education and Research (BMBF) through the FEDORA project administered by the German Aerospace Center (DLR) (01EQ2403G), and by the Carl Zeiss Foundation (P2022-00-087). This work was supported by the BMBF-funded de.NBI Cloud within the German Network for Bioinformatics Infrastructure (de.NBI) (031A532B, 031A533A, 031A533B, 031A534A, 031A535A, 031A537A, 031A537B, 031A537C, 031A537D, 031A538A).

Data collection and sharing for the Alzheimer's Disease Neuroimaging Initiative (ADNI) is funded by the National Institute on Aging (National Institutes of Health Grant U19 AG024904). The grantee organization is the Northern California Institute for Research and Education. In the past, ADNI has also received funding from the National Institute of Biomedical Imaging and Bioengineering, the Canadian Institutes of Health Research, and private sector contributions through the Foundation for the National Institutes of Health (FNIH) including generous contributions from the following: AbbVie, Alzheimer's Association; Alzheimer's Drug Discovery Foundation; Araclon Biotech; BioClinica, Inc.; Biogen; Bristol-Myers Squibb Company; CereSpir, Inc.; Cogstate; Eisai Inc.; Elan Pharmaceuticals, Inc.; Eli Lilly and Company; EuroImmun; F. Hoffmann-La Roche Ltd and its affiliated company Genentech, Inc.; Fujirebio; GE Healthcare; IXICO Ltd.; Janssen Alzheimer Immunotherapy Research & Development, LLC.; Johnson & Johnson Pharmaceutical Research &Development LLC.; Lumosity; Lundbeck; Merck & Co., Inc.; Meso Scale Diagnostics, LLC.; NeuroRx Research; Neurotrack Technologies; Novartis Pharmaceuticals Corporation; Pfizer Inc.; Piramal Imaging; Servier; Takeda Pharmaceutical Company; and Transition Therapeutics.

Data used in the preparation of this article was obtained on 17 September 2024 (imaging data) and 30 September 2024 (clinical data) from the Parkinson's Progression Markers Initiative (PPMI) database (www.ppmi-info.org/access-data-specimens/download-data), RRID:SCR_006431. For up-to-date information on the study, visit www.ppmi-info.org. PPMI – a public-private partnership – is funded by the Michael J. Fox Foundation for Parkinson's Research and funding partners, including AbbVie, Alamar Biosciences, Aligning Science Across Parkinson's (ASAP), Arrowhead Pharma, Arvinas, AskBio, BIAL, BioArctic, Biohaven, BlueRock Therapeutics, Bristol Myers Squibb, Calico Labs, Capsida Biotherapeutics, Critical Path Institute, DaCapo Brainscience, Denali, Edmond J. Safra Foundation, Eli Lilly, Gain Therapeutics, GE Healthcare, Genentech, GSK, Insitro, Johnson & Johnson Innovative Medicine, Lundbeck, Merck, Neumora, Neuron23, Novartis, Olink, Regeneron, Roche, Sanofi, Tenvie, UCB, Vanqua Bio, Voyager Therapeutics, The Weston Family Foundation.

Data were provided in part by OASIS: OASIS-3: Longitudinal Multimodal Neuroimaging: Principal Investigators: T. Benzinger, D. Marcus, J. Morris; NIH P30 AG066444, P50 AG00561, P30 NS09857781, P01 AG026276, P01 AG003991, R01 AG043434, UL1 TR000448, R01 EB009352. OASIS-4: Clinical Cohort: Principal Investigators: T. Benzinger, L. Koenig, P. LaMontagne.

Data were provided in part by the Human Connectome Project, WU-Minn Consortium (Principal Investigators: David Van Essen and Kamil Ugurbil; 1U54MH091657) funded by the 16 NIH Institutes and Centers that support the NIH Blueprint for Neuroscience Research; and by the McDonnell Center for Systems Neuroscience at Washington University. Data, methods used, and/or research reported in this publication were provided in whole or in part by the Aging Adult Vulnerability and Resiliency in the Aging Adult Brain Connectome (AABC) project (U19AG073585) and the Human Connectome Project in Aging (HCP-A, U01AG052564) funded by the National Institute of Aging of the National Institutes of Health. HCP-A was further supported by funds provided by the McDonnell Center for Neuroscience at Washington University in St. Louis. Research reported in this publication was supported by the National Institute of Mental Health of the National Institutes of Health under Award Number U01MH109589 and by funds provided by the McDonnell Center for Systems Neuroscience at Washington University in St. Louis. The HCP-Development 2.0 Release data used in this report came from DOI: 10.15154/1520708. Research using Human Connectome Project for Early Psychosis (HCP-EP) data reported in this publication was supported by the National Institute of Mental Health of the National Institutes of Health under Award Number U01MH109977. The HCP-EP 1.1 Release data used in this report came from DOI: 10.15154/1522899.

Data collection and sharing for this project was provided by the Cambridge Centre for Ageing and Neuroscience (CamCAN). CamCAN funding was provided by the UK Biotechnology and Biological Sciences Research Council (grant number BB/H008217/1), together with support from the UK Medical Research Council and University of Cambridge, UK. IXI data (Information eXtraction from Images, EPSRC GR/S21533/02) were obtained from https://brain-development.org/ixi-dataset/ under the CC BY-SA 3.0 licence.

Data were provided in part by the Brain Genomics Superstruct Project of Harvard University and the Massachusetts General Hospital, (Principal Investigators: Randy Buckner, Joshua Roffman, and Jordan Smoller), with support from the Center for Brain Science Neuroinformatics Research Group, the Athinoula A. Martinos Center for Biomedical Imaging, and the Center for Human Genetic Research. 20 individual investigators at Harvard and MGH generously contributed data to the overall project.

This manuscript was prepared using a limited access dataset obtained from the Child Mind Institute Biobank, Healthy Brain Network. This manuscript reflects the views of the authors and does not necessarily reflect the opinions or views of the Child Mind Institute.

Acknowledgements and funding statements for the other datasets used in this study are given in the Supplementary Information (Dataset acknowledgements).

A large language model was used to edit the language of the manuscript. All suggested changes were reviewed by the authors, who take full responsibility for the content of this publication.

## Tables

**Table 1. Demographic characteristics of the cohorts.**

| Cohort | N | Sites | Age, mean (SD) | Age, median [IQR] | Age range | Female, n (%) |
|---|---|---|---|---|---|---|
| Training (healthy) | 21,351 | 216 | 26.1 (16.6) | 21.8 [16.4–29.0] | 3.0–100.0 | 10,818 (50.7%) |
| Held-out test (healthy) | 2,176 | 79 | 21.2 (14.6) | 16.2 [10.7–25.9] | 5.0–79.0 | 843 (38.7%) |
| Held-out test (patients, all) | 3,684 | 64 | 20.1 (18.1) | 12.5 [9.2–23.0] | 5.0–94.0 | 1,044 (28.3%) |
| Attention-deficit/hyperactivity disorder | 1,568 | 21 | 11.8 (5.8) | 10.2 [8.4–13.1] | 5.0–50.0 | 405 (25.8%) |
| Autism | 1,074 | 39 | 16.9 (9.7) | 13.9 [10.5–20.2] | 5.2–62.0 | 146 (13.6%) |
| Major depressive disorder | 251 | 5 | 42.6 (12.1) | 42.0 [34.0–50.0] | 18.0–75.0 | 119 (47.4%) |
| Alzheimer's disease | 205 | 1 | 76.8 (6.4) | 76.0 [72.0–81.0] | 64.0–94.0 | 103 (50.2%) |
| Schizophrenia | 199 | 10 | 33.7 (11.9) | 32.0 [22.4–43.0] | 13.8–67.0 | 77 (38.7%) |
| Learning disorders | 197 | 3 | 9.6 (2.5) | 9.1 [7.8–10.8] | 5.0–17.3 | 93 (47.2%) |
| Generalized anxiety disorder | 190 | 3 | 11.7 (3.6) | 10.9 [9.0–13.9] | 5.8–21.4 | 101 (53.2%) |
| ADNI (clinical, longitudinal) | 1,229 | 62 | 73.7 (7.1) | 73.8 [69.2–78.8] | 54.4–89.6 | 566 (46.1%) |

*Diagnostic subgroups sum to the patient total (3,684 individuals; 1,044 female). Site counts for subgroups exceed the patient total of 64 because sites are shared across diagnoses. The patient subgroup is shown for completeness; it was excluded from every evaluation. ADNI ages and sex are at first visit. Trajectory groups and follow-up are given in Methods.*

## Figures

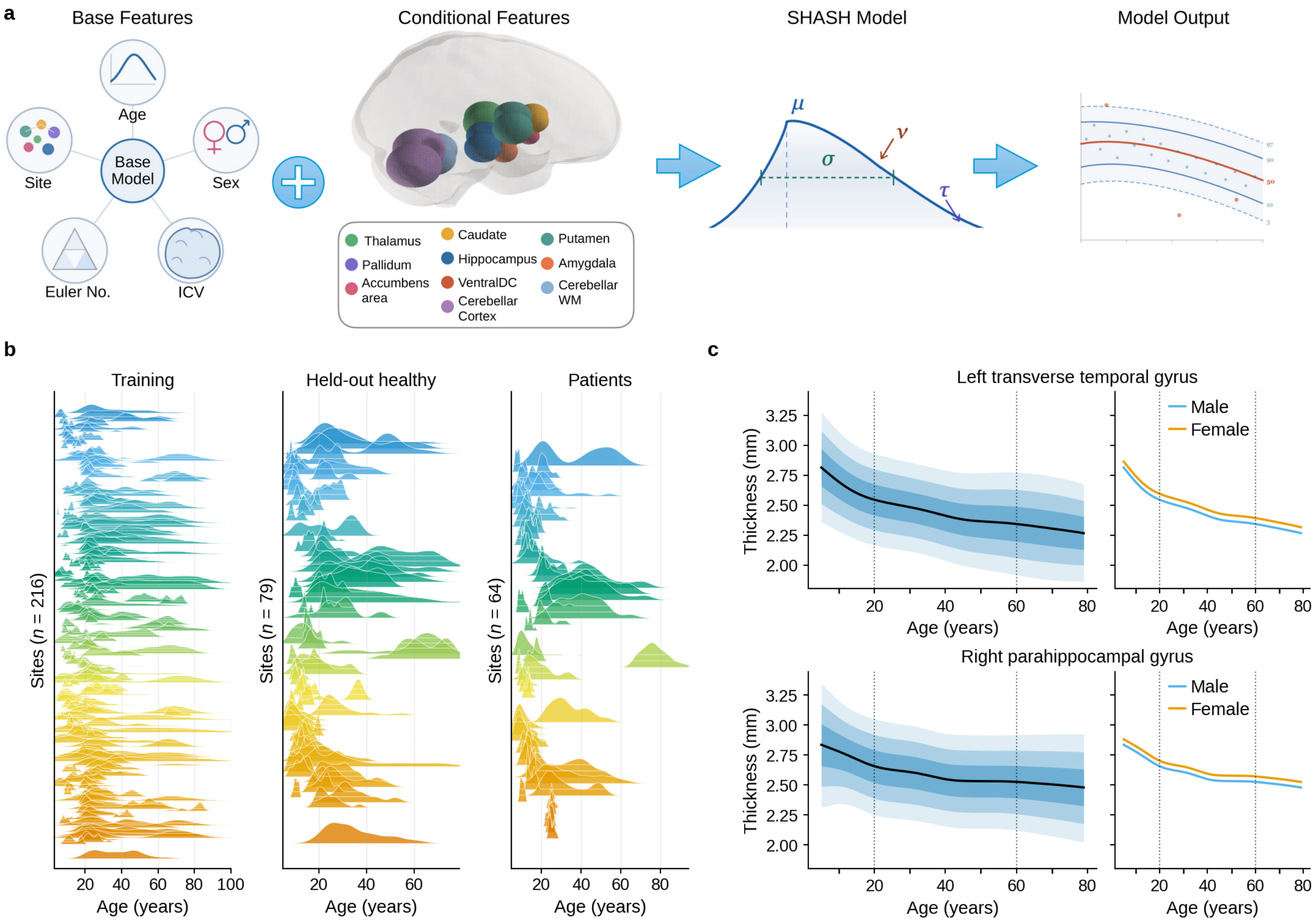

**Fig. 1.** Workflow for building conditional normative models of subcortical–cortical relationships. **a**, Schematic of the modelling framework. The base model estimates regional cortical thickness as a function of age, sex, site, Euler number, and intracranial volume. Ten conditional models are then fitted, each adding the volume of a single subcortical structure, rendered on a brain surface with the full ten-structure legend. All models are fitted as a sinh–arcsinh (SHASH) distribution, parameterized by location (μ), scale (σ), skewness (ν), and kurtosis (τ), yielding centile curves (3rd–97th) across age, overlaid on the individual observations. **b**, Ridge plots of per-site age densities for the training sample ($n$ = 21,351; 216 sites; ages 3–100), the held-out healthy test sample ($n$ = 2,176; 79 sites; ages 5–79), and held-out patient samples ($n$ = 3,684; 64 sites; ages 5–94). **c**, Normative models of two regions are shown as example, the left transverse temporal gyrus and right parahippocampal gyrus. Each is shown as a population centile fan (3rd–97th percentile, median in black) alongside the male and female median trajectories. Together the panels define the model specification, the samples on which it was fitted and evaluated, and the form of the normative fits used throughout.

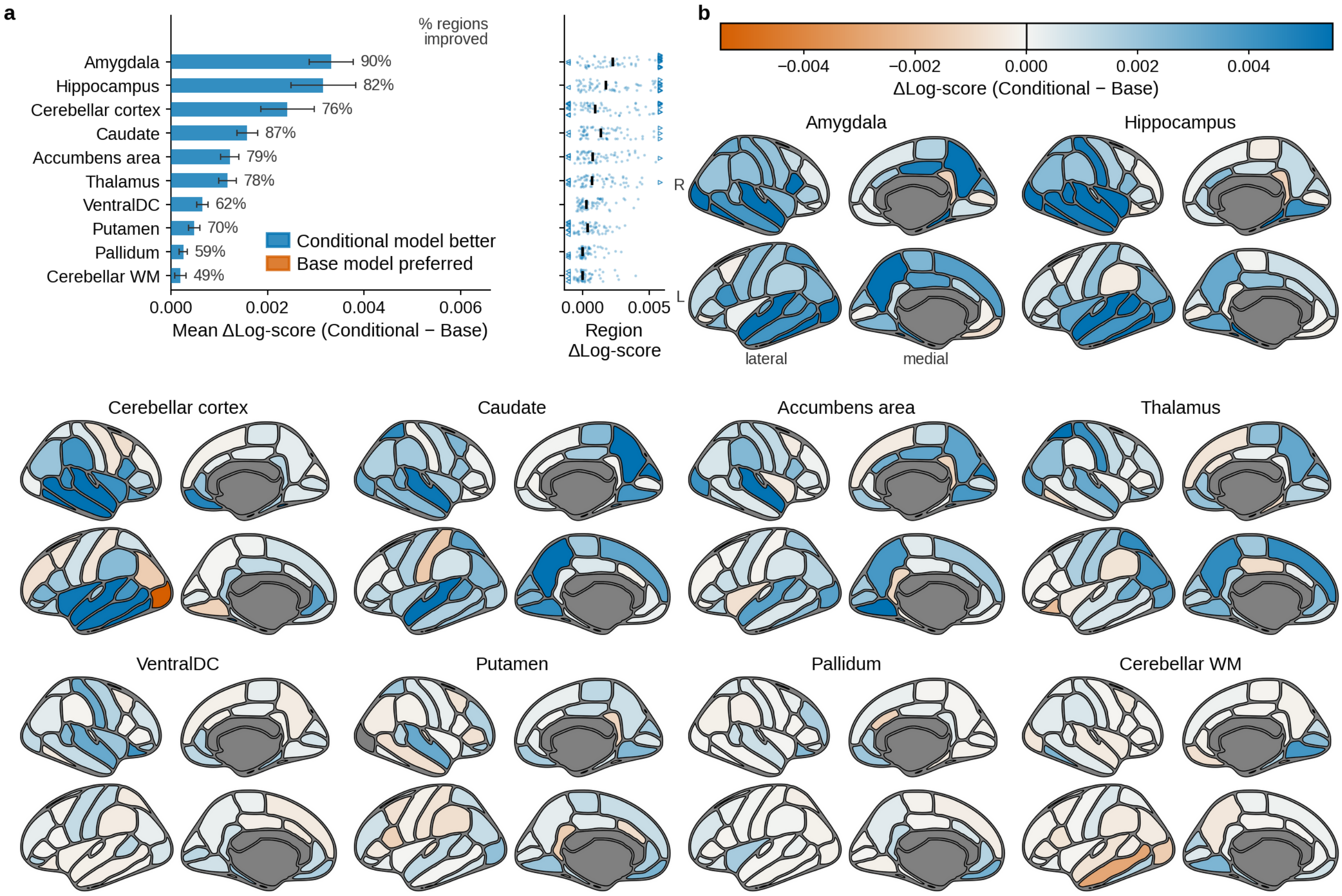

**Fig. 2.** Conditioning on subcortical volume improves out-of-sample prediction of cortical thickness on average across all ten subcortical structures in a large healthy population. **a**, Mean improvement in log-predictive-density score (ΔLog-score, conditional model minus base model) for cortical thickness, averaged across 68 Desikan–Killiany regions and evaluated on the held-out healthy sample ($n$ = 2,176), for the ten subcortical conditional models; bars are ranked by mean ΔLog-score (± SEM across regions), with the percentage of regions showing improvement annotated at right. The adjacent strip plot shows the ΔLog-score of individual regions, with the median marked by a black tick and the axis truncated at the 5th–95th percentile of the pooled values for legibility; values beyond that range (34 at each end) are drawn as triangles at the axis edge. **b**, Cortical surface maps of the region-level ΔLog-score for each of the ten conditional models (R, right hemisphere; L, left hemisphere; lateral and medial views). Cerebellar WM, cerebellar white matter; VentralDC, ventral diencephalon. Conditioning improves predictive performance on average for every structure tested, with the largest and most widespread gains for the two medial temporal lobe structures, the amygdala and hippocampus.

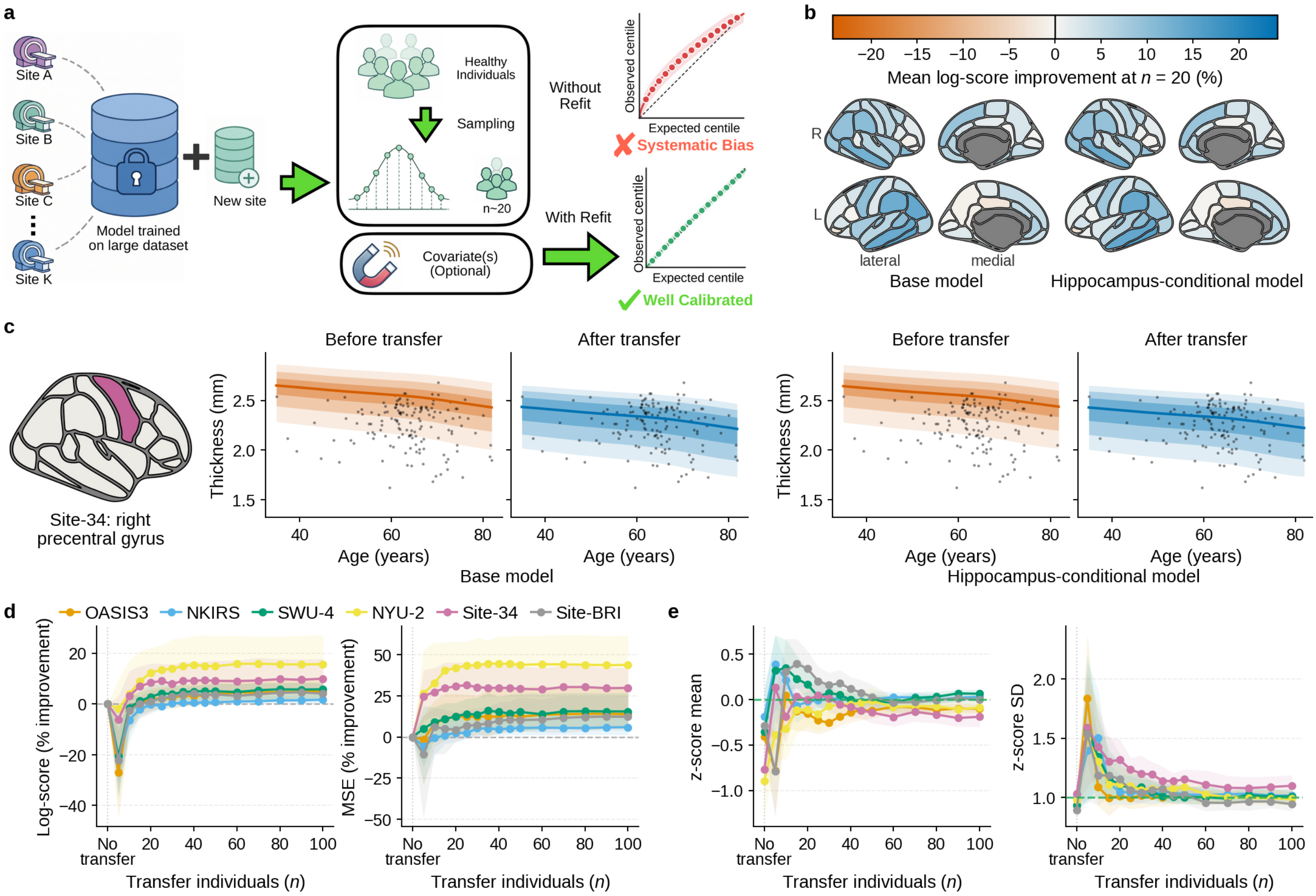

**Fig. 3.** Evaluating the calibration out-of-distribution for transfer to unseen sites. **a**, Schematic of the site-transfer procedure. A model trained across a large multi-site cohort is applied to a new site, without retraining, by drawing a small sample of healthy individuals (~20, age-stratified) from that site, optionally incorporating additional site-level covariates, and refitting an intercept-only adjustment to the model's distributional parameters. The adjusted model is then evaluated on the remaining, non-overlapping individuals at that site. Observed-versus-expected centile calibration curves illustrate that skipping this refit leaves a systematic bias, while including it yields a well-calibrated model. **b**, Whole-brain maps of the mean percentage improvement in log-score, relative to no transfer, after site transfer with 20 calibration individuals, for the base model and the hippocampus-conditional model across all cortical regions (R, right hemisphere; L, left hemisphere; lateral and medial views). **c**, Centile-fan example for the right precentral gyrus cortical thickness at one held-out site (Site-34), before and after transfer, for the base and hippocampus-conditional models. **d**, Percentage improvement in log-score and mean squared error, relative to no transfer, as a function of the number of individuals, for the base model at six held-out sites; lines are the median across the 68 cortical regions and bands the interquartile range across those regions. **e**, Mean and standard deviation of the post-calibration z-score distribution, over the same sweep and sites, against their target values of 0 and 1; lines and bands as in **d**. The transfer procedure extends offset-based recalibration[11] and is distributed as part of NormGAMLSS (Supplementary Note 7). Across sites and for both the base and conditional models, roughly 20 age-stratified individuals bring most sites close to calibration, and further calibration data adds smaller gains, with the correction acting principally on the location rather than the scale of the estimated distribution.

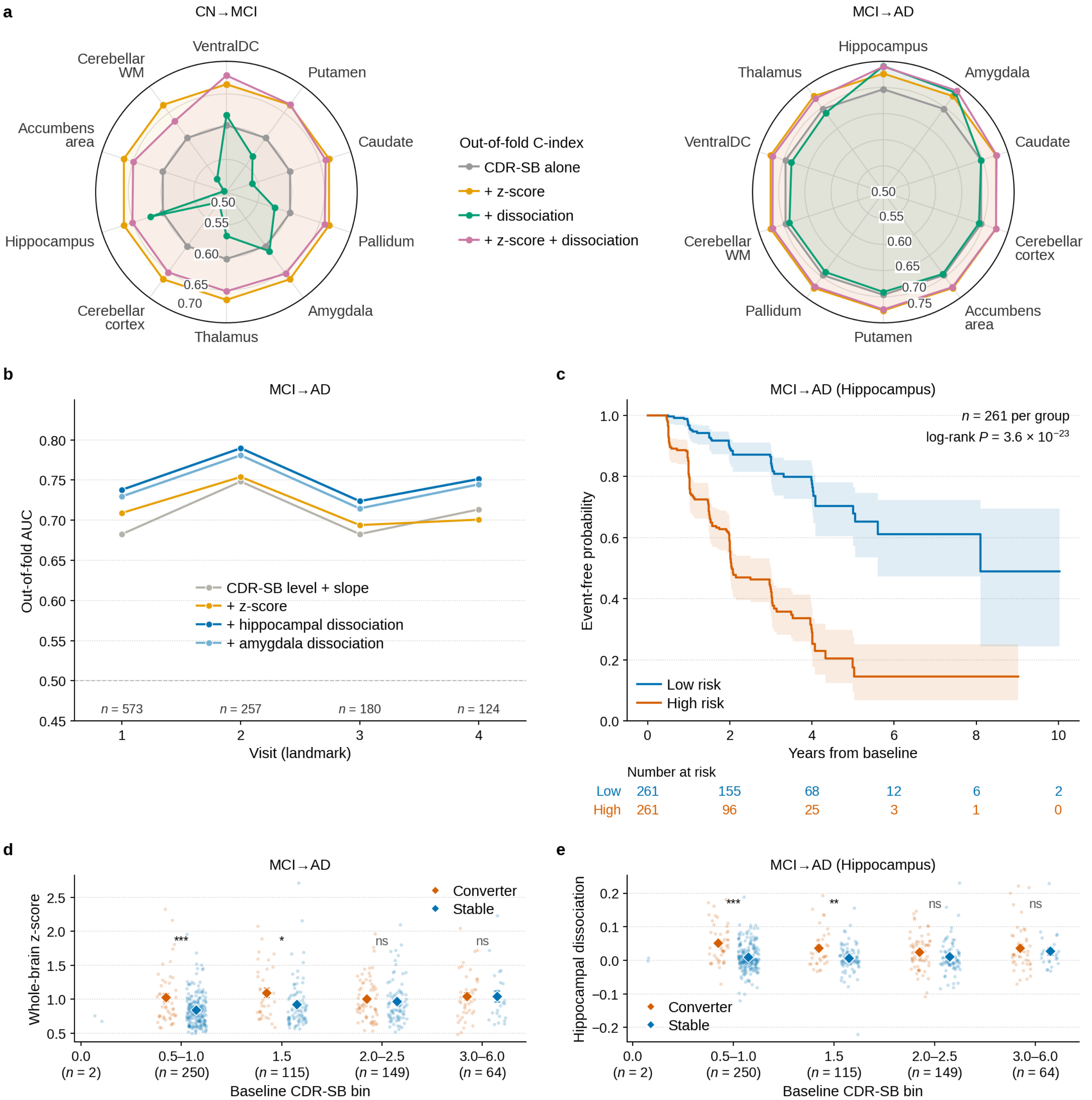

**Fig. 4.** Subcortical–cortical structural dissociation predicts conversion beyond cognitive testing, and beyond global atrophy for conversion to Alzheimer's disease dementia. ADNI was held out entirely from normative model estimation. The transfer procedure described in Fig. 3 was applied before any analysis shown here. Downstream survival and classification models were fitted within ADNI and evaluated out-of-fold by cross-validation. *Structural dissociation* denotes the reduction in whole-brain cortical deviation when the normative model is conditioned on subcortical volume. **a**, Out-of-fold concordance (Harrell's C) for Cox proportional-hazards models of time to conversion, comparing CDR-SB alone, CDR-SB + whole-brain z-score, CDR-SB + structural dissociation, and CDR-SB + both. Radar axes are the ten subcortical structures; CN→MCI (left; 184 individuals, 34 events) and MCI→AD (right; 522 individuals, 180 events). Radial axes start at 0.50. **b**, Out-of-fold discrimination (logistic-regression AUC) for eventual MCI→AD conversion from CDR-SB level and slope, and from that baseline plus whole-brain z-score, hippocampal dissociation, or amygdala dissociation. Models were evaluated at four progressively later assessment landmarks, each within its own eligible sample, whose size is given above the x axis. **c**, Kaplan–Meier event-free probability for predicted low- versus high-risk groups, split at the median of out-of-fold risk from the joint CDR-SB + z-score + hippocampal-dissociation model (522 individuals, 180

events; 261 per group), compared by log-rank test. Bands are 95% confidence intervals; numbers at risk are shown below the x axis. **d**, Whole-brain z-score in converters versus stable individuals, stratified by baseline CDR-SB severity bin, among the 580 MCI→AD individuals with a baseline CDR-SB. Asterisks denote independent two-sample *t*-tests within each bin, computed only where both groups had ≥5 individuals: * $P < 0.05$, ** $P < 0.01$, *** $P < 0.001$; ns, not significant. **e**, As in **d**, for hippocampal dissociation. Cerebellar WM, cerebellar white matter; VentralDC, ventral diencephalon. For MCI→AD, the predictive gains are largest for the hippocampus and amygdala, and baseline measures separate eventual converters, who converted a median of 1.5 years later, from stable individuals.